\documentclass[twocolumn,aps,prb]{revtex4}
\usepackage{amsmath,amssymb,mathrsfs,bm}
\usepackage{xcolor}
\usepackage{graphicx,dcolumn,times}
\begin{document}
\title{High efficiency coherent microwave-to-optics conversion via off-resonant scattering}
\author{Hai-Tao Tu$^{1,2,3}$\footnotemark[2], Kai-Yu Liao$^{1,2,3}$\footnote[2]{These authors contributed equally to this work.}\footnotemark[1], Zuan-Xian Zhang$^{1,2}$, Xiao-Hong Liu$^{1,2}$, Shun-Yuan Zheng$^{1,2}$, Shu-Zhe Yang$^{1,2}$, Xin-Ding Zhang$^{1,2,3,4}$, Hui Yan$^{1,2,3}$\footnotemark[1] and Shi-Liang Zhu$^{1,2}$\footnote[1]
{email: kaiyu.liao@m.scnu.edu.cn;yanhui@scnu.edu.cn;slzhu@scnu.edu.cn}}

\affiliation{$^1$ Guangdong Provincial Key Laboratory of Quantum Engineering and Quantum Materials, School of Physics and Telecommunication Engineering, South China Normal University, Guangzhou 510006, China}

\affiliation{$^2$ Guangdong-Hong Kong Joint Laboratory of Quantum Matter, Frontier Research Institute for Physics, South China Normal University, Guangzhou 510006, China}

\affiliation{ $^3$ GPETR Center for Quantum Precision Measurement, South China Normal University, Guangzhou 510006, China}

\affiliation{ $^4$ SCNU Qingyuan Institute of Science and Technology Innovation Co., Ltd., Qingyuan 511517, China}

\begin{abstract}
\textbf{ Quantum transducers that can convert quantum signals from the microwave to the optical domain are a crucial optical interface for quantum information technology. Coherent microwave-to-optics conversions have been realized with various physical platforms, but all of them are limited to low efficiencies of less than 50\%, the threshold of the no-cloning quantum regime. Here we report a coherent microwave-to-optics transduction using Rydberg atoms and off-resonant scattering technique with an efficiency of $82\pm 2\%$ and a bandwidth of about 1 MHz. The high conversion efficiency is maintained for microwave photons range from thousands to about 50, suggesting that our transduction is readily applicable to the single-photon level. Without requiring cavities or aggressive cooling to quantum ground states, our results would push atomic transducers closer to practical applications in quantum technologies.}
\end{abstract}
\maketitle

Coherent and efficient transduction of microwave into optical light plays a critical role in developing  quantum technologies. For instance, in order to build a large-scale quantum network with superconducting quantum computers \cite{KimbleNature2008, WehnerSci2018}, we need a transducer capable of transducing gigahertz microwave into terahertz optical light, which offers  low transmission loss in room-temperature environments\cite{OBrienNP2009}. Moreover, the optical domain provides access to a suit of well-developed quantum optical tools, including highly efficient single photon detectors and long-lived quantum memories \cite{LvovskyNP2009}. Notably, the access to single photon detector would facilitate the detection and imaging of weak microwave signals with potential applications in astronomy, medicine, and other fields. To realize these applications, a coherent microwave-to-optics transducer with near-unity efficiency and large bandwidths is essential \cite{XiangRMP2013, LambertReview2019, LaukReview2020}.

Compared to classical transducers, a high-efficiency quantum transducer plays a more crucial role in preserving the fragile quantum state of information. For example, the quantum capacity of a bosonic channel is finite only if its photon transmissivity is greater than 50\% \cite{WolfPRL2007}, and thus a conversion efficiency above this threshold will be necessary for any applications of transferring quantum states within the no-cloning regime without post-selection \cite{GrosshansPRA2001}.
Several promising platforms have been proposed and implemented to realize the microwave up-conversion, including electro-optics \cite{RuedaOptica2016, FanSA2018, WitmerQST2020}, optomechanics \cite{BochmannNP2013, AndrewsNP2014, HigginbothamNP2018, ForschNP2019, WuPRAppl2020, JiangNC2020}, optically active dopants in solids \cite{WilliamsonPRL2014, OBrienPRL2014, HisatomiPRB2016, WelinskiPRL2019, GonzalvoPRA2019, BartholomewNC2020}, and cold atoms\cite{KiffnerNJP2016, WenhuiPRL2018, SaffmanPRA2019}. Of these approaches, the highest efficiencies  attained to date are 47\% through optomechanics\cite{HigginbothamNP2018} and 25\% through electro-optics\cite{FanSA2018}. Both  need high-quality cavities to achieve strong nonlinearities \cite{LambertReview2019}, which inevitably limits the conversion bandwidth. Therefore, achieving near-unity microwave up-conversion compatible with large bandwidths remains an outstanding challenge.

Cold atomic systems provide a natural setting for the realization of hybrid quantum interfaces. Cold atomic ensembles exhibit excellent cooperativity along phase-matched direction and large nonlinearity to achieve the single-photon level transduction\cite{ZibrovPRL2002}, so optical cavities are not crucial for achieving high efficiency \cite{PetrosyanNJP2019}.  In addition, cold atoms coupled with superconducting resonators can store microwave photons in the long-lived hyperfine levels, and it is possible to combine the transduction of optical photons with quantum memories that will lay the basis of quantum network nodes\cite{PetrosyanPRA2009}. Recently, microwave-to-optical transduction through a cold atomic ensemble has been demonstrated with a 5\% conversion efficiency \cite{WenhuiPRA2019}. However, the maximally achievable efficiency  is limited since the atoms will evolve into a coherent population-trapped dark state and are decoupled from the microwave or optical fields\cite{HarrisPRL2000}.

\begin{figure*}[htp]
\begin{center}
\includegraphics[width=17.5cm]{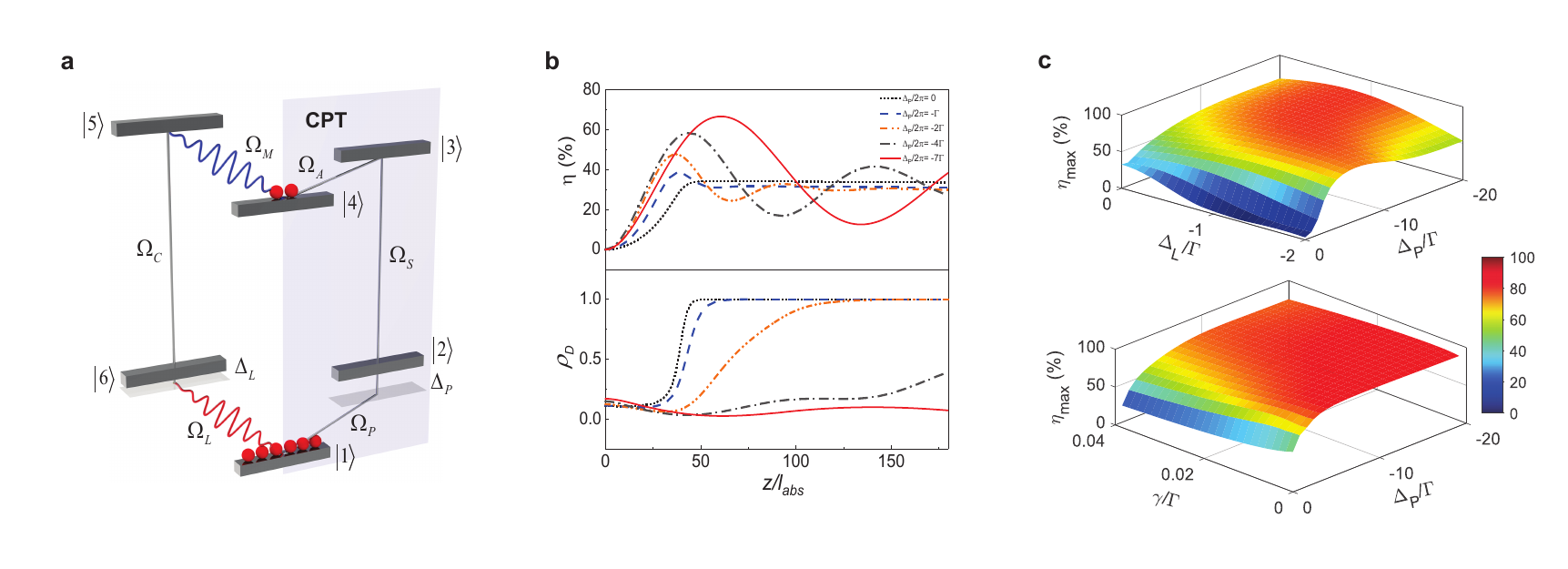}
\caption{\label{fig:theory} \textbf{Theoretical analysis of microwave-to-optics conversion via off-resonant scattering}.
\textbf{a}, Energy level configuration. Level $|1\rangle$ is a ground state. Levels $|2\rangle$ and $|6\rangle$ are two low-lying states with decay rates $\Gamma$ and $\Gamma^{\prime}$, respectively. Levels $|3\rangle$, $|4\rangle$, and $|5\rangle$ are three Rydberg states with dephasing rate $\gamma$. The microwave field ($\Omega_\mathrm{M}$, blue wavy) is converted into an optical field ($\Omega_\mathrm{L}$, red wavy) through frequency-mixing process. $\Omega_\mathrm{X}$ ($\mathrm{X} \in \{$P, S, A, C$\}$, grey straight) are four auxiliary fields. $\Delta_\mathrm{P}$ and $\Delta_\mathrm{L}$ are the single photon detunings of $\Omega_\mathrm{P}$ and $\Omega_\mathrm{L}$ respectively. 
\textbf{b}, Conversion efficiency $\eta$ (top panel) versus the propagation distance $z$ for different $\Delta_\mathrm{P}$ ($\Delta_\mathrm{L}$ = 0 $\Gamma$ and $\gamma$ = 0.03 $\Gamma$). Dark state probability $\mathcal{P}_{\mathrm{D}}$ (bottom panel) versus the propagation distance $z$ for different $\Delta_\mathrm{P}$.
\textbf{c}, Maximum efficiency $\eta_\mathrm{max}$ (top panel) as a function of $\Delta_\mathrm{L}$ and $\Delta_\mathrm{P}$ for a given dephasing rate $\gamma$ = 0.03 $\Gamma$. Bottom panel shows $\eta_\mathrm{max}$ as a function of $\gamma$ and $\Delta_\mathrm{P}$ for optimized conversion with detuning $\Delta_\mathrm{L}$ = -0.8 $\Gamma$. In \textbf{b} and \textbf{c}, the theoretical results are obtained from the numerical integration of Maxwell-Bloch equations under the steady-state condition, with parameters of the $^{87}$Rb atoms $\{\Omega_\mathrm{P}=0.3 \Gamma, \Omega_\mathrm{S}=1.5 \Gamma, \Omega_\mathrm{A}=0.2 \Gamma, \Omega_\mathrm{M}=0.001 \Gamma, \Omega_\mathrm{C}=2 \Gamma, \gamma^{\prime}=0.002 \Gamma, \Gamma^{\prime}=0.17 \Gamma, b=28.7\}$.}
\end{center}
\end{figure*}

Here, we solve the difficulties faced by previous transduction experiments that use neutral atoms \cite{WenhuiPRL2018, WenhuiPRA2019} by  creating a large atomic coherence  with auxiliary fields to enhance the absorption of microwave photons. We do this by using the off-resonant six-wave mixing technique and a long, thin cylindrical atomic-gas medium. Our transducer demonstrates an unprecedented photon-conversion efficiency of $82\pm 2\%$, which is significantly higher than the quantum no-cloning limit, and maintains a large bandwidth of around 1 MHz at efficiencies above 50\%. Moreover, our converter features  excellent preservation of phase information during the conversion process, which is confirmed through optical heterodyne measurement, allowing for faithful transduction of single-photon quantum states from the microwave to the optical domain.

\bigskip
\noindent\textbf{Results}\\
\textbf{The off-resonant six-wave mixing scheme.}
We begin with a brief description of the six-wave mixing  scheme for microwave-to-optics conversion, with a six-level atomic system as shown in Fig.~\ref{fig:theory}a. In this scheme, the transduction of the input microwave field $\Omega_\mathrm{M}$ into the optical field $\Omega_\mathrm{L}$ is achieved through six-wave mixing assisted by four driven fields $\Omega_\mathrm{X}$ ($\mathrm{X} \in \{$P, S, A, C$\}$. The dynamics of the atomic system is governed by the following master equation for the density operator $\varrho$:
\begin{eqnarray}
\begin{array}{ll}
&\partial_{t} \varrho=-\frac{i}{\hbar}\left[H, \varrho\right]+\mathcal{L}_{\Gamma}\varrho+\mathcal{L}_{\text{deph}}\varrho,
\end{array}
\end{eqnarray}
where $H$ is the Hamiltonian for an independent atom interacting with six external fields, and the terms $\mathcal{L}_{\Gamma}\varrho$ and $\mathcal{L}_{\text{deph}}\varrho$ represent the spontaneous emission of low-lying states and the dephasing of atomic coherence involving the Rydberg states, respectively. In the paraxial approximation, we assume that the depletion of the fields $\Omega_\mathrm{S}$ and $\Omega_\mathrm{C}$ is small in the medium and that the other four fields are treated in a self-consistent approach (more details see Supplemental Section 1).


\begin{figure*}[ptb]
\begin{center}
\includegraphics[width=18cm]{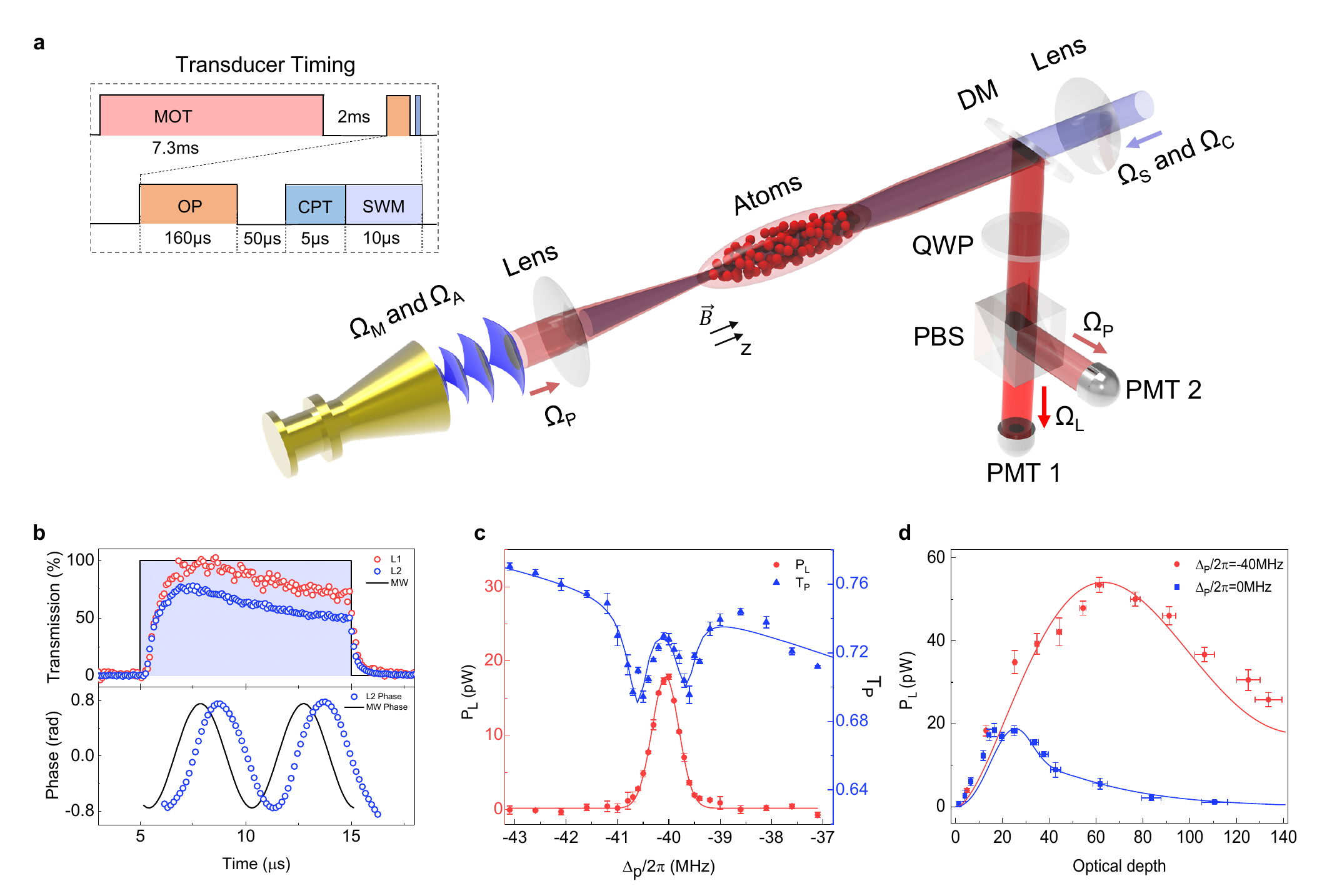}
\caption{\label{fig:sys} \textbf{ Experimental demonstration of microwave-to-optics conversion in cold atoms}.
\textbf{a}, Experimental set-up and time sequence. Six-wave mixing (SWM) is performed by collinearly propagating fields along the $z$ axis in a cigar-shaped atomic cloud. In each experimental cycle, the atomic coherence is established by auxiliary fields $\Omega_\mathrm{P}$, $\Omega_\mathrm{S}$, and $\Omega_\mathrm{A}$  after  switching off the MOT and optical pumping (OP). A microwave pulse $\Omega_\mathrm{M}$ ($\approx$37 GHz) emitting from a circularly polarized antenna scatters off the coherence with field $\Omega_\mathrm{C}$, generating an optical pulse $\Omega_\mathrm{L}$ ($\approx$780 nm). The copropagating fields $\Omega_\mathrm{L}$ and $\Omega_\mathrm{P}$ are separated using a polarizing beam-splitter (QWP + PBS) and measured by the photomultiplier tubes (PMT1 and PMT2), respectively.
\textbf{b}, Temporal waveforms of the input microwave pulse (black) and output optical pulses at $\Omega_\mathrm{M}/2\pi$ = 0.2 (0.5) MHz denoted as  $L_{1}$ ($L_{2}$). The ``SWM'' block in \textbf{a} corresponds to the square pulse in \textbf{b}.  Transmission is scaled by the optical power converted with 100\% efficiency. The bottom panel depicts the relative phase of a heterodyne signal for the phase-modulated microwave. The solid line is the input sine modulation of 100 kHz and 0.8 amplitude, and blue circles are experimental data extracted from numerical phase detection. \textbf{c}, Spectra of the transmission $T_\mathrm{P}$ (blue) and of the generated optical power $P_\mathrm{L}$ (red) at $\Omega_\mathrm{M}/2\pi$ = 0.1 MHz. $T_\mathrm{P}$ is fitted to $\exp(-\alpha L)$, with the absorption coefficient $\alpha\propto N\operatorname{Im}(\varrho_{\mathrm{P}})$ and the atomic polarization $\varrho_{\mathrm{P}}$ of the microwave-dressed four-level system. The spectrum of $P_\mathrm{L}$ is fitted to a Lorentz function. \textbf{d}, $P_\mathrm{L}$ versus optical depth for off-resonant (red) and
near-resonant (blue) scatterings at $\Omega_\mathrm{M}/2\pi$ = 0.16 MHz. The symbols represent experimental data, and the solid lines are theoretically simulated curves. The fitting parameters $\{\Omega_\mathrm{P}, \Omega_\mathrm{S}, \Omega_\mathrm{A}, \Omega_\mathrm{C}, \Gamma, \gamma_{3}, \gamma_{4}, \gamma_{5}, \gamma^{\prime}_{3},\gamma^{\prime}_{4}, \gamma^{\prime}_{5}, \Gamma^{\prime}\}$ are $2\pi\times \{2.1, 8.3, 0.94, 12.5, 6, 0.3, 0.08, 0.1, 0.01, 0.01, 0.01, 1\}$ MHz, respectively. The error bars in \textbf{c} and \textbf{d} indicate 1$\sigma$ standard error from three measurements. }
\end{center}
\end{figure*}

The frequency-mixing process is simulated through the numerical integration of Maxwell-Bloch equations under the steady-state condition. Here, an ensemble of $^{87}$Rb atoms is under consideration. Figure~\ref{fig:theory}b shows the calculated conversion efficiency $\eta$ (top panel) as a function of the propagation distance $z$ for several detunings $\Delta_\mathrm{P}$. The distinct behaviors of different detunings $\Delta_\mathrm{P}$ in the spatial evolution of scattered fields can be explained by the probability $\mathcal{P}_{\mathrm{D}}$  (bottom panel) in the dark state $|D\rangle$  located at a distance $z$, where $|D\rangle \propto\left(\Omega_\mathrm{M}^{*} \Omega_\mathrm{S}^{*}|1\rangle-\Omega_\mathrm{M}^{*} \Omega_\mathrm{P}|3\rangle+\Omega_\mathrm{A}^{*} \Omega_\mathrm{P}|5\rangle\right)$.
For near-resonant scattering ($|\Delta_\mathrm{P}|\leq \Gamma$), the atomic population in $|D\rangle $ increases with the buildup of photon conversion along the propagation direction. The input and output fields evolve without further interaction with the medium, and thus $\eta$ saturates when almost all of the atoms are trapped in the dark state\cite{WenhuiPRL2018}.


For off-resonant scattering ($|\Delta_\mathrm{P}|\gg \Gamma$), a two-photon transition consisting of the detuned fields $\Omega_\mathrm{P}$ and $\Omega_\mathrm{S}$ establishes an effective coupling on the $|1\rangle\leftrightarrow|3\rangle$ transition. In the absence of the microwave field $\Omega_\mathrm{M}$, the effective field and an auxiliary field $\Omega_\mathrm{A}$ create quantum coherence between the ground state $|1\rangle$ and Rydberg state $|4\rangle$ through coherent population trapping (CPT). The large atomic coherence $\varrho_{14}$ significantly prevents the system from being trapped in $|D\rangle $. As illustrated in the bottom panel of Fig~\ref{fig:theory}b, $\mathcal{P}_{\mathrm{D}}$ is maintained below 30\% within more than a hundred of the absorption lengths ($l_{abs}=\Gamma^{\prime}/4\zeta_{\mathrm{L}}$, where $\Gamma^{\prime}$ is the decay rate of state $|6\rangle$, and $\zeta_{\mathrm{L}}$ is the coupling constant at the $|1\rangle\leftrightarrow|6\rangle$ transition). Thus, the energy oscillates back and forth between the input microwave field and the up-converted optical field, making it possible to choose the propagation distance to maximize  conversion efficiency \cite{JainPRL1996}, as shown in  the top panel of Fig~\ref{fig:theory}b.

Furthermore, we investigate the optimal parameters for improving the efficiency of off-resonant scattering. We extract the maximum efficiency $\eta_\mathrm{max}$ from the spatial evolution of scattered fields within over a hundred absorption lengths. Figure~\ref{fig:theory}c shows the maximum conversion efficiency $\eta_\mathrm{max}$ as a function of the laser detunings $\Delta_\mathrm{P}$ and $\Delta_\mathrm{L}$ as well as the Rydberg state dephasing rate $\gamma$, respectively. 
A small detuning ($|\Delta_\mathrm{L}| \approx \Gamma$), which lessens the re-absorption of the converted optical field inside the atomic ensemble, can increase conversion efficiency. 
With an optimized detuning $\Delta_\mathrm{L}$, the conversion efficiency $\eta_\mathrm{max}$ increases with the decreasing of the dephasing rate $\gamma$. 
$\eta_\mathrm{max} \approx$ 0.85 can be achieved with the parameters $\{\Delta_\mathrm{P}, \Delta_\mathrm{L}, \gamma, \mathrm{OD}\}$ of $\{-18 \Gamma, -0.8 \Gamma, 0.02 \Gamma, 120\}$. Note that increasing  $\Delta_\mathrm{P}$ leads to further improvement because the residual occupation of state $|2\rangle$ is reduced in the six-level system, but this would require a much thicker medium to reach the maximum efficiency (see Supplementary Section 2 for more details).


\begin{table*}
\begin{tabular}{p{1.4cm}<{\centering}|p{2.2cm}<{\centering}|p{2.2cm}<{\centering}|p{2.0cm}<{\centering}|p{2.0cm}<{\centering}|p{2.0cm}<{\centering}|p{3cm}<{\centering}}
\hline Field & $f$ or $\lambda$  & Transition & Polarization  & Waist ($\mu$m) & $|\boldsymbol{d}_{l m}|$ ($ea_{0}$) & Peak Rabi frequency\\
\hline
$\Omega_{\mathrm{P}}$ & $\sim$ 384.2 THz & $|1\rangle\leftrightarrow|2\rangle$ & $\sigma^{+}$ & 56(3) & 2.99  & 2$\pi$ $\times$ 3.1(2) MHz\\
$\Omega_{\mathrm{L}}$ & (780.2 nm) & $|1\rangle\leftrightarrow|6\rangle$ &  $\sigma^{-}$ & NA & 1.22 & NA \\
\hline
$\Omega_{\mathrm{S}}$ & $\sim$ 623.5 THz  & $|2\rangle\leftrightarrow|3\rangle$ & $\sigma^{-}$ & 54(3) & 0.006 & 2$\pi$ $\times$ 9.2(4) MHz \\
$\Omega_{\mathrm{C}}$ & (480.8 nm)  & $|5\rangle\leftrightarrow|6\rangle$ & $\sigma^{+}$ & 54(3) & 0.013  & 2$\pi$ $\times$ 13.2(6) MHz\\
\hline
$\Omega_{\mathrm{A}}$ & 36.705 GHz  & $|3\rangle\leftrightarrow|4\rangle$ & $\sigma^{+}$ & NA & 363.6  & 2$\pi$ $\times$ 0.94(1) MHz\\
$\Omega_{\mathrm{M}}$ & 36.907 GHz  & $|4\rangle\leftrightarrow|5\rangle$  & $\sigma^{+}$ & NA & 667.3  & NA \\
\hline
\end{tabular}
\caption{\label{tbl:fields} \textbf{Six external fields and electric transition dipole moments.} The Table lists for each field used in our experiment the respective frequency \cite{GallagherBook1994}, polarization, waist, the dipole moment $|\boldsymbol{d}_{l m}|$ for relevant electric transition and the estimated peak Rabi frequency.
}
\end{table*}

\bigskip
\noindent\textbf{Experimental setup.}
We implement our microwave-to-optical transducer with an ensemble of cold $^{87}$Rb atoms released from a two-dimensional magneto-optical trap (2D-MOT) in order to obtain a large optical depth (OD)\cite{WangNP2019, LiaoPRA2020}. Figure~\ref{fig:sys}a shows a simplified diagram of the experimental setup: full details are included in the Methods. The fields $\Omega_\mathrm{P}$, $\Omega_\mathrm{A}$, and $\Omega_\mathrm{M}$  collinearly propagate along the $z$ axis and counterpropagate with the fields $\Omega_\mathrm{S}$ and $\Omega_\mathrm{C}$, which are derived from a single 480 nm laser. The microwave fields $\Omega_\mathrm{A}$ and $\Omega_\mathrm{M}$, from two microwave generators (R\&S SMF100A), are combined with a power combiner just before the circular-polarization horn antenna. All of the input lasers are focused onto the front end of the cigar-shaped atomic ensemble to increase the conversion area of microwave and mitigate the polarization shift along with the Gaussian beam propagation. 
The fields $\Omega_\mathrm{P}$ and $\Omega_\mathrm{L}$, with opposite circular polarization, are separated by a quarter-wave plate (QWP) and a polarization beam splitter (PBS). The detail parameters of six fields and relevant electric dipole moments are presented in Table 1. The intensities of $\Omega_\mathrm{L}$ and $\Omega_\mathrm{P}$ are measured with two photomultiplier tubes (PMT, Hamamatsu, H10720-20). A laser line bandpass filter is placed in front of PMT to separate the stray noise photons, and the average time-dependent signals are recorded by a 5 GHz high-speed digital scope (R\&S RTE1024).

The whole experiment is periodically run with a repeat rate of 100 Hz, which consists of 7.3 ms MOT loading time followed by a 2.7 ms conversion window (see upper left of figure 2a). 
During the conversion window, a bias magnetic field of 6.4 G along the $z$ axis is added. The laser-cooled atoms are optically pumped to a specific Zeeman state $|5S_{1/2}, F=2, m_F=2\rangle$. Subsequently, a $T$= 10 $\mu$s, microwave pulse $\Omega_\mathrm{M}$ is emitted into the atomic ensemble for up-conversion with a 5 $\mu$s delay after the auxiliary fields are switching on, where the delay guarantees the atomic coherence between states $|1\rangle$ and $|4\rangle$ is established.

\bigskip
\noindent\textbf{Coherent microwave up-conversion.}
We carry out the experiment following the theoretical simulation. The auxiliary fields $\Omega_\mathrm{S}$ and $\Omega_\mathrm{C}$ are blue detuned from the corresponding atomic transition by 40~MHz and 4.8~MHz, respectively \cite{SibalicCPC2017}. As shown in Figure~\ref{fig:sys}b (top panel), the input square-modulated microwave pulse has a rise and fall time of around 5 ns. Firstly, we scan the detuning $\Delta_\mathrm{P}$ of $\Omega_\mathrm{P}$ across the $|1\rangle\leftrightarrow|3\rangle$ two-photon resonance and  simultaneously measure the transmission of the field $\Omega_\mathrm{P}$ and the power of the converted field $\Omega_\mathrm{L}$ (Fig.~\ref{fig:sys}c). The transmission $T_{P}$ exhibits a double-peak shape primarily resulting from the effect of the microwave-dressed electromagnetically induced absorption (EIA) \cite{LiaoPRA2020}. The spectrum of the converted field ($P_\mathrm{L}$) features a pronounced peak around $\Delta_\mathrm{P}$ = -40~MHz. After that, we vary the medium length by blocking the trapping laser beam of MOT with an aperture, while maintaining the lasers focused on the ensemble fore-end. Figure~\ref{fig:sys}d (red dots) corroborates that the converted fileld ($P_\mathrm{L}$) reaches a maximum at an OD around 63. For comparison, we also carry out the experiment in the near-resonant scheme\cite{WenhuiPRA2019}. The maximally converted field decreases to only one third of that of the off-resonant scheme. The converted field starts to attenuate at larger ODs, which is mainly due to the re-absorption of the resonant field $\Omega_\mathrm{L}$ (see Supplementary Section 5). The solid curves in Fig.~\ref{fig:sys}d represent theoretical simulations, which are in good agreement with the experiment.

As shown in Figure~\ref{fig:sys}b (top panel), the generated laser field $\Omega_\mathrm{L}$ exhibits an exponential decay profile in the falling edge. The decay time is about 0.18 $\mu$s, which agrees with the lifetime of the low-lying state $|6\rangle$ ($\tau = 1/\Gamma^{\prime}$). To verify the coherence property of this transducer, we modulate the phase of the pulse $\Omega_\mathrm{M}$ with a Sine function and then recover the phase modulation with optical heterodyne detection\cite{WenhuiPRL2018}. The heterodyne measurement is performed between the converted signal $\Omega_\mathrm{L}$ and a reference beam for a pulse duration of 100 $\mu$s, which is derived from the same laser as $\Omega_\mathrm{P}$. The frequency difference between the two beating laser beams is 2 MHz. Figure~\ref{fig:sys}b (bottom panel) shows that the phase information is almost perfectly transferred during the conversion with an average fidelity of 98\%, confirming the phase-preserving nature of our convertor. The observed delay between the recovery and the input of phase modulation is mainly derived from the slow-light effects in the frequency-mixing process \cite{FleischhauerRMP2005}.

\begin{figure*}[ptb]
\begin{center}
\includegraphics[width=16cm]{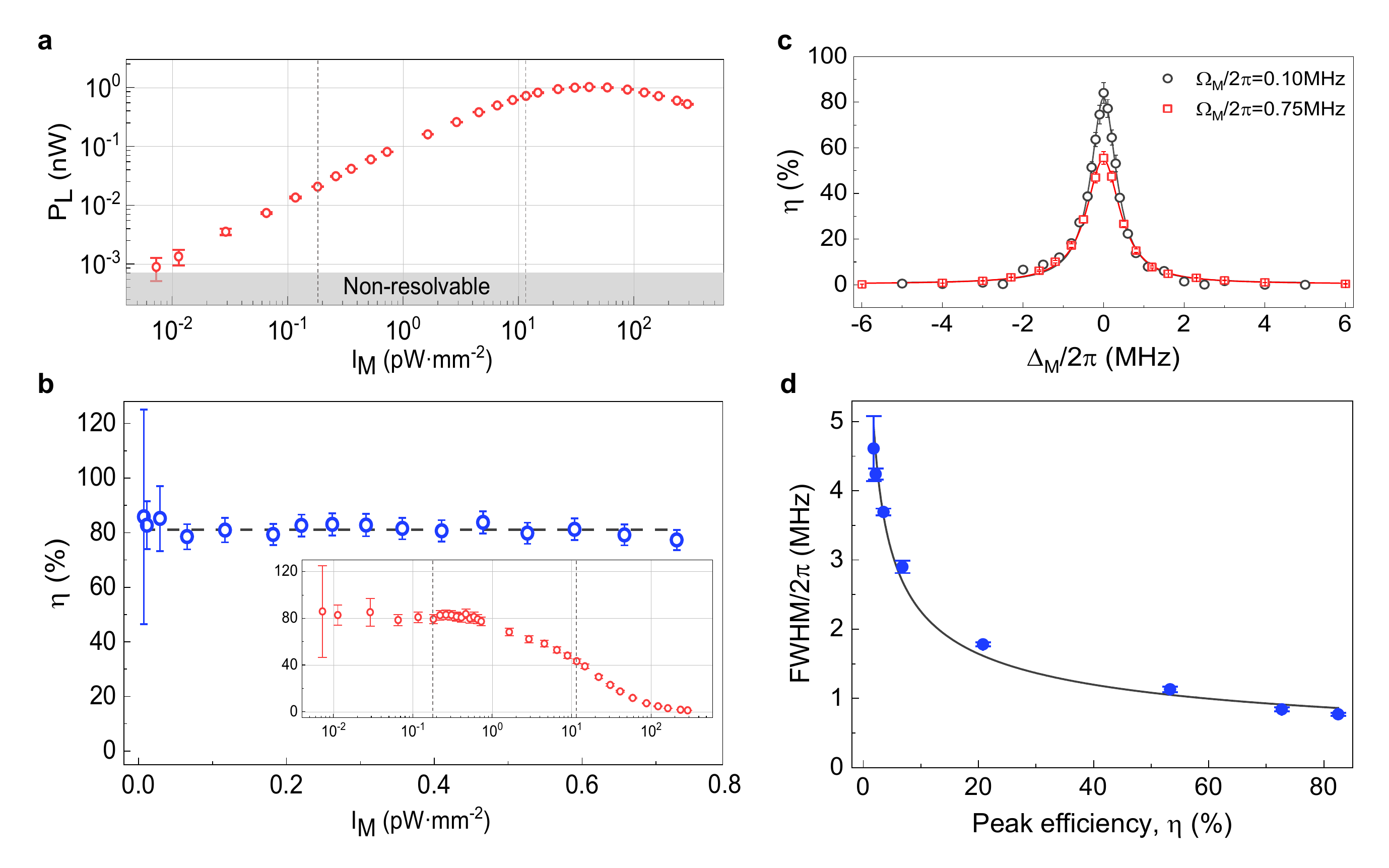}
\caption{\label{data}  \textbf{Conversion efficiency and bandwidth via off-resonant scattering}.
\textbf{a}, Light power $P_\mathrm{L}$ versus input microwave intensity $I_\mathrm{M}$. The grey region corresponds to the experimentally non-resolvable power signals after averaging over 10,000 scans. \textbf{b}, Efficiency $\eta$ calculated for the linear region of conversion shown in \textbf{a}. The dashed line indicates the average value of efficiency. The inset depicts $\eta$ against the entire data of $I_\mathrm{M}$ in \textbf{a}. \textbf{c}, Two sample spectra of $\eta$ versus microwave detuning $\Delta_\mathrm{M}$. Their microwave intensities are indicated with the vertical dashed lines in \textbf{a} and \textbf{b}. \textbf{d}, FWHM versus peak efficiency $\eta$. FWHM is extracted by fitting a Lorentz function to the microwave spectrum at different intensities $I_\mathrm{M}$. Solid line is the result of polynomial fitting. Error bars in \textbf{a} and \textbf{d} indicate 1$\sigma$ standard error from three measurements, and the error bars in \textbf{b} and \textbf{c} are calculated through the error propagation with the optical power error bars shown in \textbf{a}, the 1.8\% microwave intensity uncertainty, and the 4.7\% averaged cross section uncertainty.}
\end{center}
\end{figure*}

\bigskip
\noindent\textbf{Efficiency and bandwidth.}
As shown in Figure~\ref{data}a, the converted power $P_\mathrm{L}$ grows approximately linearly in the weak-field regime ($\Omega_\mathrm{M}\ll \Omega_\mathrm{A}$). The linearity starts to break down when $\Omega_\mathrm{M}\approx \Omega_\mathrm{A}$. Then, the converted power $P_\mathrm{L}$ drops at higher microwave intensities ($\Omega_\mathrm{M}\gg \Omega_\mathrm{A}$). The photon conversion efficiency of our setup is calculated as
\begin{eqnarray}
\begin{array}{ll}
&\eta=\frac{P_\mathrm{L} / \hbar \omega_\mathrm{L}}{I_\mathrm{M} S_\mathrm{M} / \hbar \omega_\mathrm{M}}.
\end{array}
\end{eqnarray}
We only consider the microwave photon incident to the conversion region, where the atomic ensemble and all six fields overlap and $S_\mathrm{M}$ is the averaged cross section of the conversion medium. The conversion efficiency calculation is consistent with its definition based on the photon fluxes. In Fig~\ref{data}b, we obtain an average conversion efficiency of $\eta$ = 82\%, with a standard deviation 2\% and an average uncertainty 7\%. The input microwave pulse contains a mean photon number ranging from 50 to 6400, calculated by $I_\mathrm{M} S_\mathrm{M} T/ \hbar \omega_\mathrm{M}$.

To analyze the conversion bandwidth, we measure the efficiency as a function of the detuning of the microwave field. Two sample microwave spectra and their fits to the Lorentz function are depicted in Fig~\ref{data}c. The full width at half maximum (FWHM) is $\approx$ 0.77 MHz at the peak efficiency of about 82\% (grey circles). Fig~\ref{data}d shows the FWHM as a function of the peak efficiency in the microwave spectrum. The spectrum has a FWHM of around 1 MHz in the weak-field regime. Due to power broadening, the FWHM grows close to 4.7 MHz with a higher microwave intensity. The conversion bandwidth in our study is several to dozens of times larger than the values reported using electro-optics\cite{FanSA2018}, rare-earth ions ensemble\cite{HisatomiPRB2016, GonzalvoPRA2019} and optomechanics\cite{HigginbothamNP2018} approaches, illustrating the broadband conversion ability of atomic transducers.

\bigskip
\noindent\textbf{Discussion}\\
Our atomic transducer operates with the continuous-wave auxiliary fields in a steady-state atomic polarization. An all-resonant six-wave mixing approach was experimentally realized in Refs. \cite{WenhuiPRL2018, WenhuiPRA2019}, where a conversion efficiency around $5\%$ was reported. Compared with these experiments, we made several improvements to achieve high conversion efficiency. We overcome a main drawback for all-resonant approach by realizing a simplified  off-resonant scheme theoretically proposed in Ref \cite{KiffnerNJP2016}: the conversion efficiency saturates in all-resonant approach because the distribution in the dark state is large and even approaches to one under the condition of a large OD. The dark state decouples from all external fields, so the absorbtion of the microwave $\Omega_M$ is small when the distribution in the dark state is large. In contrast,  we prepare a significant atomic coherence between the metastable states $|1\rangle$ and $|4\rangle$ with the off-resonant auxiliary fields and
thus a high conversion efficiency is achievable because the distribution in the dark state is negligible even in a large OD system, as demonstrated in Fig. 1b.
To achieve a high-efficiency conversion atomic transducer, large ODs and low Rydberg-state dephasing rates are two necessary conditions. Unlike the previous transducer with a dense atomic cloud of small size, our large ODs are achieved by implementing the 2D-MOT configuration with a cylindrical trapping volume\cite{WangNP2019}, thereby maintaining a relatively low density to minimize the imperfections due to Rydberg-Rydberg interactions. Besides the low atomic density, the Rydberg-state dephasings are suppressed by reducing the stray magnetic fields with three pairs of Helmholtz coils, as well as by lifting the Zeeman degeneracy with a stable bias magnetic field.

At present, as shown in Fig~\ref{data}a, the resolvable microwave pulse containing around fifty photons is primarily limited by the background noise fluctuation of the PMT detector.  In principle, with the low-noise single photon counters and better optical filters, the microwave-to-optics conversion can be performed at the single-photon level. In Supplemental Materials, we estimate the number of optical photons converted by the microwave background and it is about 0.8 at temperature $300\ K$. So the effect of the thermal photons is negligible in the most cases and contributes a fluctuation of about $2\%$ in the worst case in our experiments. The thermal microwave reduces with the decrease of temperature, for instance, the number of the converted photons becomes 0.1 at temperature $39\ K$ and thus the effect of the microwave background can be also neglected in a single-photon experiment when temperature is lower than $39\ K$. Moreover, a larger Rabi frequency $\Omega_\mathrm{C}$ can further improve the conversion bandwidth in our scheme\cite{WenhuiPRA2019}.

%

In summary, we have realized an atomic transducer with high efficiency via off-resonant six-wave mixing in free space. With large ODs and low Rydberg-state dephasing rates, we have obtained a coherent microwave-to-optics conversion efficiency of $82\pm 2\%$ and bandwidth of about 1 MHz; the conversion efficiency exceeds the 50\% threshold value for practical applications.
The  scheme may have various potential applications. Besides neutral atoms based transduction, the off-resonant scattering scheme can be used to increase the efficiency of rare-earth crystal based transduction, such as three-wave mixing using spin transitions for the ground states \cite{WilliamsonPRL2014, OBrienPRL2014} or optically excited states\cite{WelinskiPRL2019}. The microwave-to-optics conversion approach can be further developed to perform at the single-photon level. Combined with high-efficient single photon detectors, the scheme can be used in detecting and imaging weak microwave signals, which has various applications in astronomy, medicine, and other fields.

Notably, the high efficiency and broad bandwidth achieved in our scheme meet the optical interface requirement for superconducting qubits.
As pioneering works that coherent coupling between Rydberg atoms and the microwave field of various superconducting devices have been achieved\cite{HoganPRL2012, AviglianoPRA2014, MorganPRL2020, FortaghARXIV2021}, we can expect atomic transducers to convert single-photon quantum states in a cryogenic environment\cite{BagciNature2014, SaffmanPRA2017, MirhosseiniNature2020}, providing a critical component of large-scale hybrid quantum networks. For instance, this transduction approach can be used in an ensemble of atoms coherently coupled to the microwave field of an on-chip coplanar waveguide resonator, resulting in a directional emission of optical photons even without an optical cavity \cite{PetrosyanNJP2019}. The microwave photon residence in cavity increases the interaction time of frequency-mixing, and it relaxes the medium thickness demand compared to the free-space conversion. Moreover, this Rydberg transducer operates in a steady-state atomic polarization associated with the Rydberg level coupled to the microwave field (i.e., $|4\rangle$ in our scheme), leading to a collective enhancement of microwave transition in the single-photon regime.  Meanwhile, the resonant dipole-dipole interaction between Rydberg atoms becomes negligible in the case of single microwave photon. Therefore, the atomic transducer developed here may find practical applications in large-scale hybrid quantum networks.


\bigskip
\noindent\textbf{Methods}\\
{
\noindent\textbf{Conversion efficiency.}
The conversion efficiency $\eta$ is defined as the ratio of the converted optical photon flux to the input microwave photon flux. In the theoretical analysis of figure 1, the efficiency is given by:
\begin{eqnarray}
\begin{array}{ll}
&\eta=\frac{b |\Omega_{\mathrm{L}}^{\mathrm{out}}|^{2} }{ |\Omega_{\mathrm{M}}^{\mathrm{in}}|^{2} },
\end{array}
\end{eqnarray}
where $b$ represents the ratio of the coupling constants $\zeta_{\mathrm{M}}$ to $\zeta_{\mathrm{L}}$. $\Omega_{\mathrm{M}}^{\mathrm{in}}$ and $\Omega_{\mathrm{L}}^{\mathrm{out}}$ are the Rabi frequencies of the input microwave field and converted optical field, respectively.

\bigskip
\noindent\textbf{Dark state.}
The approximate dark state of the Hamiltonian in Eq. (1) is of the form $|D\rangle = \mathcal{C} \left(\Omega_{\mathrm{M}}^{*} \Omega_{\mathrm{S}}^{*}|1\rangle-\Omega_{\mathrm{M}}^{*} \Omega_{\mathrm{P}}|3\rangle+\Omega_{\mathrm{A}}^{*} \Omega_{\mathrm{P}}|5\rangle\right)$  for $\Omega_\mathrm{L} / \Omega_\mathrm{P}=-\Omega_\mathrm{A}^{*} \Omega_\mathrm{C}^{*} /\left(\Omega_\mathrm{M}^{*} \Omega_\mathrm{S}^{*}\right)$, where the normalisation constant is
\begin{eqnarray}
\begin{array}{ll}
&\mathcal{C}=\frac{1}{\sqrt{\Omega_{\mathrm{A}}^{2} \Omega_{\mathrm{P}}^{2}+\Omega_{\mathrm{M}}^{2}\left(\Omega_{\mathrm{P}}^{2}+\Omega_{\mathrm{S}}^{2}\right)}}.
\end{array}
\end{eqnarray}
The probability for the atoms populated in the dark state $|D\rangle$ is defined as:
\begin{eqnarray}
\begin{array}{ll}
&\mathcal{P}_{\mathrm{D}}=\operatorname{Tr}\left[\varrho_{\mathrm{D}}(z) \varrho(z)\right],
\end{array}
\end{eqnarray}
where $\varrho$ is the steady-state solution of Eq. (1) at the given position $z$.

\bigskip
\noindent\textbf{Cold atom preparation and the Rydberg-laser system.}
The cold $^{87}$Rb medium with a typical size of 4 $\times$ 4 $\times$ 24 mm$^{3}$ serves as the microwave-to-optical transducer. The relevant energy levels are $|1\rangle=|5S_{1/2}, F=2, m_F=2\rangle$, $|2\rangle = |5P_{3/2}, F=3, m_F=3\rangle$, $|3\rangle=|39D_{3/2}, m_{J}=1/2\rangle$, $|4\rangle=|40P_{3/2}, m_{J}=-1/2\rangle$, $|5\rangle=|39D_{5 / 2}, m_{J}=1/2\rangle$, and $|6\rangle=|5P_{3/2},F=2, m_F=1\rangle$, respectively. For the two-dimensional magneto-optical trap, the typical OD is 140 for the transition $|5S_{1/2}, F=2, m_F=2\rangle\leftrightarrow|5P_{3/2}, F=3, m_F=3\rangle$, which yields an atomic number density of around 1.2 $\times$ 10$^{\rm{10}}$ cm$^{-3}$. Each trapping laser beam has a power of 25 mW and a radius of 1.6 cm. The total power of the two repump laser beams is about 35 mW with the same radius as the trapping laser beam. The gradient of the quadruple magnetic field is 8 G cm$^{-1}$, and the temperature of the atomic cloud is about 200 $\mu$K. In each operation cycle, the MOT quadruple gradient magnetic field is switched off before the 2.7 ms conversion  window. Subsequently, a bias magnetic field of 6.4 G along the longitudinal direction of the medium is switched on to define the quantization axis, and then the atoms in molasses are optically pumped into the ground state $|5S_{1/2}, F=2, m_F=2\rangle$.

The 780 nm laser and the seed of the 480 nm laser are frequency locked to a high finesse ultra-stable Fabry-Perot cavity by the Pound-Drever-Hall technique \cite{LiaoPRA2020}. The linewidth of the 780 nm laser is estimated to be 2~kHz and the linewidth of the 480 nm laser is below 4~kHz. The optical beams for the auxiliary fields $\Omega_{\mathrm{P}}$, $\Omega_{\mathrm{S}}$, and $\Omega_{\mathrm{C}}$ are focused on the front end of the atomic cloud with the $1/e^2$ radii $\textit{w}_{\mathrm{P}}$, $\textit{w}_{\mathrm{S}}$, and $\textit{w}_{\mathrm{C}}$ of 56(3), 54(3), and 54(3)~$\mu$m, respectively. For the optimized conversion using a 21.5 mm long medium, the beam radii at the rear end of the atomic cloud are 105(3), 79(1), and 79(1)~$\mu$m, respectively. The peak Rabi frequencies of the auxiliary fields are $\Omega^{(0)}_{\mathrm{P}}/2\pi$ = 3.1(2) MHz, $\Omega^{(0)}_{\mathrm{S}}/2\pi$ = 9.2(4) MHz, $\Omega^{(0)}_{\mathrm{A}}/2\pi$ = 0.94(1) MHz, and $\Omega^{(0)}_{\mathrm{C}}/2\pi$ = 13.2(6) MHz, respectively.


\bigskip
\noindent\textbf{Experimental parameter calibration.}
To determine the internal conversion efficiency of our transducer, both the microwave intensity $I_{\mathrm{M}}$ and averaged receiving cross section $S_{\mathrm{M}}$ should be accurately calibrated. The microwave intensity $I_{\mathrm{M}}$ at the medium is self-calibrated through the four-level EIA scheme\cite{GordonAPL2014, LiaoPRA2020} (more details see Supplemental Section 4). The microwave field amplitude is first measured in the Autler-Townes splitting regime to calibrate the power generater and then the calibrated power meter is used to extrapolate the microwave field at low intensities. The receiving cross section $S_{\mathrm{M}}$ for a cigar-shaped cloud is estimated by calculating the average of absorption cross section in the axial direction (see Supplemental Section 3). By using the two parameters calibrated as above and the simulated conversion efficiency, the theoretical calculation of converted optical powers agrees well with the experimental results as shown in Fig. 2(d).

In addition, the optical depth is derived from the fits of two-level transmission spectra, attained when only the field $\Omega_{\mathrm{P}}$ is incident on the ensemble. The Rabi frequencies of the laser beams ($\Omega_{\mathrm{P}}$, $\Omega_{\mathrm{S}}$, and $\Omega_{\mathrm{C}}$) are calibrated by measuring the respective laser intensities \cite{SibalicCPC2017}. Moreover, the Rydberg dephasing rates $\gamma_{3}$, $\gamma_{4}$, and $\gamma_{5} $ are extracted from the fit with four-level EIA spectra as described above.

\bigskip
\noindent\textbf{Phase information transfer fidelity.}
The classical fidelity of coherent microwave-to-optics conversion is defined as:
\begin{eqnarray}
\begin{array}{ll}
&\frac{|\int e^{i\varphi_{\mathrm{M}}(t-t_{d})} e^{-i\varphi_{\mathrm{L}}(t)}dt|^{2}}{[\int|e^{-i\varphi_{\mathrm{M}}(t)}|^{2} dt] [\int|e^{-i\varphi_{\mathrm{L}}(t)}|^{2} dt]},
\end{array}
\end{eqnarray}
where $\varphi_{\mathrm{M}}(t)$ ($\varphi_{\mathrm{L}}(t)$) represents the input (recovered) phase modulation waveform in the optical heterodyne measurements, and $t_{d}$ is the delay time.

\bigskip
\noindent\textbf{Data availability}

The main data supporting the results in this study are available within the paper and its Supplementary Information. The raw datasets  are available from the corresponding authors upon reasonable request.

\bigskip
\noindent\textbf{Code availability}

The codes used for the theoretical simulations  are available from the corresponding authors upon reasonable request.


\bigskip
\bigskip\noindent\textbf{Acknowledgements}\\
\noindent We thank Thibault Vogt, Wenhui Li, Jingshan Han, and Weibin Li for useful discussions. The work was supported by  the Key-Area Research and Development Program of GuangDong Province (Grants No. 2019B030330001 and No. 2020B0301030008), the National Natural Science Foundation of China (Grants No. 91636218, No. 11822403, No. 11804104,  No. 61875060, No. U1801661, and No. U20A2074), the Key Project of Science and Technology of Guangzhou (Grant No. 2019050001), the National Key Research and Development Program of China (Grant No. 2020YFA0309500), and the Natural Science Foundation of Guangdong Province (Grants No.2018A030313342 and No. 2018A0303130066).\\

\bigskip
\noindent\textbf{Author Contributions}\\
\noindent K.Y.L. and H.Y. designed the experiment.
H.T.T., Z.X.Z, X.H.L. and S.Z.Y. carried out the experiments.
H.T.T., K.Y.L., S.Y.Z. and X.D.Z. conducted raw data analysis.
K.Y.L., H.Y. and S.L.Z. wrote the paper, and all authors discussed the paper contents.
H.Y. and S.L.Z. supervised the project.

\bigskip
\noindent\textbf{Competing Financial Interests}\\
\noindent The authors declare no competing financial interests.

\newpage

\renewcommand{\theequation}{S\arabic{equation}}
\setcounter{equation}{0}
\renewcommand{\thefigure}{S\arabic{figure}}
\setcounter{figure}{0}
\maketitle

\section*{SI. Maxwell-Bloch equations}
The numerical simulation for microwave-to-optical conversion in Rydberg atoms is performed within the standard framework of the Maxwell-Bloch equations from first principals. Owing to the low atomic density and small population of Rydberg states, the dipole-dipole interactions between Rydberg atoms are negligible and are sketched by effective dephasing rates in the independent-atom picture. The time evolution of the single-atom density operator $\varrho$ is described by a Markovian master equation,
\begin{equation}
\partial_{t} \varrho=-\frac{i}{\hbar}\left[H, \varrho\right]+\mathcal{L}_{\Gamma}\varrho+\mathcal{L}_{\text{deph}}\varrho.
\end{equation}
In electric-dipole and rotating-wave approximations, the Hamiltonian for a single atom interacting with six external fields is given by
\begin{equation}
\label{H}
\begin{aligned}
H=&-\hbar \left(\Delta_\mathrm{P} \sigma_{2 2}+\Delta_\mathrm{L} \sigma_{66} \right)-\hbar \sum_{k=3}^{5}\Delta_{k} \sigma_{kk}\\
&-\hbar\left(\Omega_{\mathrm{P}} \sigma_{21}+\Omega_{\mathrm{S}} \sigma_{32}+\Omega_{\mathrm{A}} \sigma_{34}\right.\\
&\quad\quad \left.+\Omega_{\mathrm{M}} \sigma_{54}+\Omega_{\mathrm{C}} \sigma_{56}+\Omega_{\mathrm{L}} \sigma_{61}+\mathrm{H.c.}\right),
\end{aligned}
\end{equation}
where $\sigma_{i j}=|i\rangle\langle j|$ are atomic transition operators. The detuning $\Delta_{k}$ in Eq. (\ref{H}) is defined as
\begin{subequations}
\begin{align}
\Delta_\mathrm{P} &=\omega_{\mathrm{P}}-\omega_{2},\\
\Delta_\mathrm{L} &=\omega_{\mathrm{L}}-\omega_{6}, \\
\Delta_{3}&=\omega_{\mathrm{P}}+\omega_{\mathrm{S}}-\omega_{3}, \\
\Delta_{4}&=\omega_{\mathrm{P}}+\omega_{\mathrm{S}}-\omega_{\mathrm{A}}-\omega_{4}, \\
\Delta_{5}&=\omega_{\mathrm{P}}+\omega_{\mathrm{S}}+\omega_{\mathrm{M}}-\omega_{\mathrm{A}}-\omega_{5},
\end{align}
\end{subequations}
where $\hbar\omega_{k}$ is the energy of level $|k\rangle$ with respect to $|1\rangle$, and $\omega_{\mathrm{X}}$ is the frequency of field $\Omega_{\mathrm{X}}$ ($X \in \{\mathrm{P}, \mathrm{S}, \mathrm{A}, \mathrm{M}, \mathrm{C}, \mathrm{L} \}$) with the wave vector $\boldsymbol{k}_{\mathrm{X}}$. The Rabi frequency of each field is
\begin{equation}
\Omega_{\mathrm{X}}=\frac{\boldsymbol{d}_{l m} \cdot \boldsymbol{e}_{\mathrm{X}}}{\hbar} \mathcal{E}_{\mathrm{X}},
\end{equation}
and the matrix element of the electric dipole moment operator $\hat{\boldsymbol{d}}$ on the transition $|l\rangle\leftrightarrow|m\rangle$ is
\begin{equation}
\boldsymbol{d}_{l m}=\langle l|\hat{\boldsymbol{d}}| m\rangle,
\end{equation}
where $\boldsymbol{e}_{\mathrm{X}}$ is the unit polarization vector and $\mathcal{E}_{X}$ is the complex amplitude of field $\Omega_{\mathrm{X}}$. The six laser and microwave fields form a closed loop,
\begin{equation}
\omega_{\mathrm{P}}+\omega_{\mathrm{S}}+\omega_{\mathrm{M}}-\omega_{\mathrm{A}}-\omega_{\mathrm{C}}-\omega_{\mathrm{L}} = 0,
\end{equation}
and the phase matching condition
\begin{equation}
\boldsymbol{k}_{\mathrm{P}}+\boldsymbol{k}_{\mathrm{S}}+\boldsymbol{k}_{\mathrm{M}}-\boldsymbol{k}_{\mathrm{A}}-\boldsymbol{k}_{\mathrm{C}}-\boldsymbol{k}_{\mathrm{L}}= 0
\end{equation}
is met. The term $\mathcal{L}_{\Gamma}\varrho$ in Eq. (S1) describes the spontaneous emission, and writes
\begin{equation}
\begin{aligned}
\mathcal{L}_{\Gamma} \varrho=&-\frac{\Gamma}{2}\left(\sigma_{22} \varrho+\varrho \sigma_{22}-2 \sigma_{12} \varrho \sigma_{12}^{\dagger}\right) \\
&-\frac{\gamma^{\prime}_{3}}{2}\left(\sigma_{33} \varrho+\varrho \sigma_{33}-2 \sigma_{23} \varrho \sigma_{23}^{\dagger}\right) \\
&-\frac{\gamma^{\prime}_{4}}{2}\left(\sigma_{44} \varrho+\varrho \sigma_{44}-2 \sigma_{14} \varrho \sigma_{14}^{\dagger}\right)\\
&-\frac{\gamma^{\prime}_{5}}{2}\left(\sigma_{55} \varrho+\varrho \sigma_{55}-2 \sigma_{65} \varrho \sigma_{65}^{\dagger}\right) \\
&-\frac{\Gamma^{\prime}}{2}\left(\sigma_{66} \varrho+\varrho \sigma_{66}-2 \sigma_{16} \varrho \sigma_{16}^{\dagger}\right) ,
\end{aligned}
\end{equation}
where $\Gamma$ and $\Gamma^{\prime}$ are the decay rates of excited state $|2\rangle$ and $|6\rangle$ respectively \cite{DuPRA2009}, and the terms proportional to $\gamma^{\prime}_{3}$, $\gamma^{\prime}_{4}$, $\gamma^{\prime}_{5}$ in Eq. (S8) account for the decay of the long-lived Rydberg states $|3\rangle$, $|4\rangle$ and $|5\rangle$. In our scheme, we can neglect the Rydberg decay terms since they are much smaller than the decay rate of the excited states.

The last term in Eq. (S1) presents the dephasing of atomic coherence owing to the Rydberg-Rydberg interaction, atomic collision, finite laser linewidth and stray electromagnetic field, giving
\begin{equation}
\mathcal{L}_{\text{deph}} \varrho=-\sum_{k=3}^{5}\gamma_{k}\left(\sigma_{kk} \varrho+\varrho \sigma_{kk}-2 \sigma_{kk} \varrho \sigma_{kk}^{\dagger}\right).
\end{equation}
To facilitate the theoretical analysis, the decay (dephasing) rates of the three Rydberg states are treated as the same value of $\gamma^{\prime}$ ($\gamma$) in Fig.~1 of the main text. In the experimental simulation, the dephasing rates $\gamma_{k}$ are free parameters and are determined by the spectra fitting to experimental data.

The strong auxiliary fields $\Omega_{\mathrm{S}}$ and $\Omega_{\mathrm{C}}$ change insignificantly in the conversoin process, and we take into account all the other fields $\Omega_{\mathrm{Y}}$ ($\mathrm{Y}\in \{\mathrm{P}, \mathrm{A}, \mathrm{M}, \mathrm{L} \}$) for the self-consistent Maxwell-Bloch equations. In the slowly varying envelope and paraxial approximation, the propagation of field $\Omega_{\mathrm{Y}}$ is governed by
\begin{equation}
\left(\frac{1}{c} \partial_{t}+\partial_{z}\right) \Omega_{\mathrm{Y}}=2 \mathrm{i} \zeta_{\mathrm{Y}} \varrho_{ji}.
\end{equation}
The coupling constant $\zeta_{\mathrm{Y}}$ is given by
\begin{equation}
\zeta_{\mathrm{Y}}=\frac{\mathcal{N}\left|\boldsymbol{d}_{ji}\right|^{2}}{2 \hbar \epsilon_{0} c} \omega_{\mathrm{Y}},
\end{equation}
where $\mathcal{N}$ is the average atomic density and $c$ is the speed of light. The set of Eq. (S1) and Eq. (S10) stands for a system of the coupled partial differential equations, and the Maxwell-Bloch equations are numerically solved in the steady-state condition by using MATLAB software.

\section*{SII. Optical-depth requirement for optimized conversion}
\begin{figure}[h]
\begin{center}
\includegraphics[width=8cm]{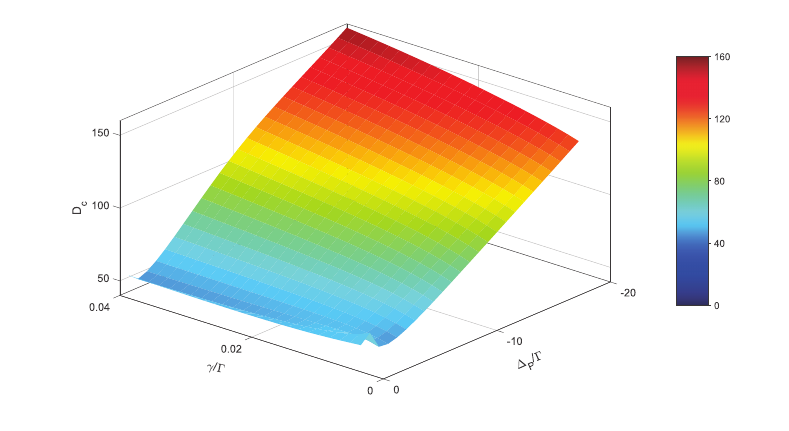}
\caption{ \textbf{ Optical depth $D_{c}$ as a function of $\gamma$ and $\Delta_\mathrm{P}$.}  The optical depth $D_{c}$ required for maximum efficiency is extracted from the simulation result of spatial evolution, and the parameters are the same as that in the bottom panel of Fig. 1(c) in main text. }
\end{center}
\end{figure}

Here we analyze the dependence of the optical depth $D_{c}$ on the auxiliary field detuning and the Rydberg-state dephasing rate. Since the interaction strength between atoms and photons is decreased in the off-resonant scattering, one must increase the medium thickness to reach the maximum conversion efficiency with respect to the resonant six-wave mixing. As shown in Fig. S1, $D_{c}$ required for the optimized conversion with laser detuning $\Delta_\mathrm{P}$ = -20 $\Gamma$ is about triple that of the resonant scattering case. Note that the increase of atomic number density can give rise to the non-negligible Rydberg-Rydberg interactions, which prevent part of the atoms from involving in the frequency-mixing process. The interaction-induced imperfection results in the absorption of converted field, and thus decrease the conversion effieciency \cite{Peyronel2012}. In principal, a moderate optical depth is necessary to achieve the near-unity conversion efficiency with the Rydberg ensemble in a small size.

\section*{SIII. Averaged cross section of conversion medium}
\begin{figure}[h]
\begin{center}
\includegraphics[width=7cm]{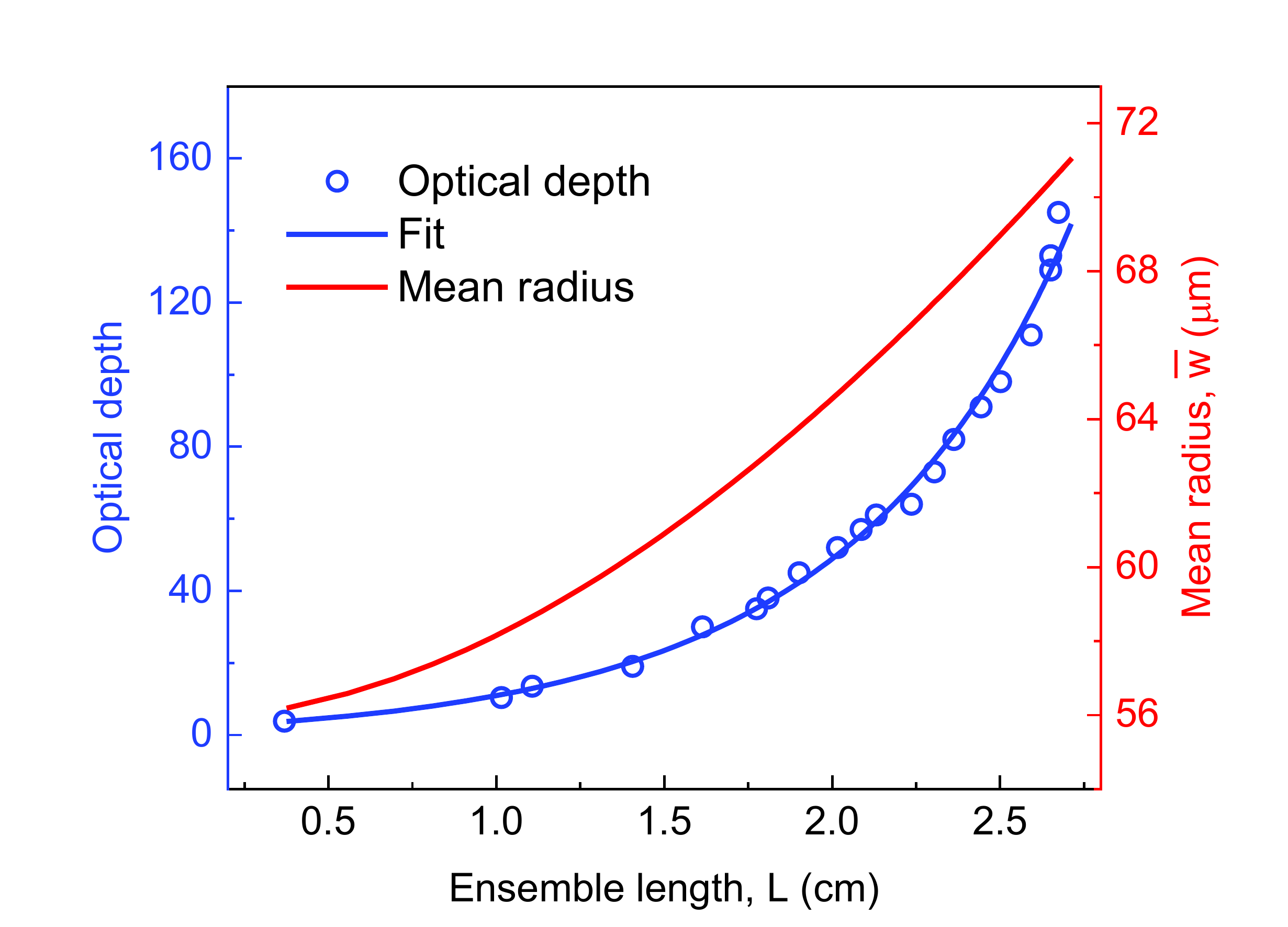}
\caption{\label{radius} \textbf{ Optical depth and mean radius versus the longitudinal length of atomic ensemble. } Longitudinal length is measured by using the geometry image method. The data is fit to a polynomial function. The mean radius $\overline{\textit{w}}$ against the different medium length is calculated using Eq. S(15).
 }
\end{center}
\end{figure}

In our system, the laser beams for auxiliary fields $\Omega_{\mathrm{P}}$, $\Omega_{\mathrm{S}}$, and $\Omega_{\mathrm{C}}$ are focused on front end of the cigar-shaped atomic cloud with the $1/e^2$ radii $\textit{w}_{\mathrm{P}}$, $\textit{w}_{\mathrm{S}}$, and $\textit{w}_{\mathrm{C}}$ of 56, 54, and 54~$\mu$m, respectively. Notably, the microwave up-conversion merely happens in the blue beam volume where all the auxiliary lasers overlap in the atomic cloud. The relative position between the beam waist and the medium is rigorously located by CCD.
Since the radius of the cylindric ensemble is much bigger than that of the laser beams, the atomic density is almost homogeneous in the radial direction of the interaction volume. In the experiment, the ensemble length is comparable to the Rayleigh range of the blue laser beam. To estimate the effective receiving cross section of conversion medium, we therefore consider both the Gaussian atomic density profile and the blue beam's expansion in $z$ direction as follows.

We model the atomic density profile in its longitudinal direction by
\begin{equation}
\tilde{n}(z)=n_{\mathrm{max}} e^{-2\left[(z-L/2)/ \textit{w}\right]^{2}} \theta\left(z\right) \theta\left(L-z\right),
\end{equation}
where $\theta\left(z\right)$ denotes the Heaviside step function, and $\textit{w}$ = $2L/3$ is the $1/e^2$ half width of the Gaussian density profile. For a given optical depth OD and ensemble length L, the peak atomic density $n_{\mathrm{max}} $ in Eq. (S12) is written as
\begin{equation}
n_{\mathrm{max}}=\frac{{\mathrm{OD}} \Gamma}{ \bar{\beta}}\left(\int_{0}^{L} e^{-2\left[(z-L/2)/ \textit{w}\right]^{2}}  \mathrm{~d} z\right)^{-1},
\end{equation}
where $\bar{\beta}=2\omega_{\mathrm{P}}\left|\boldsymbol{d}_{21}\right|^{2} /\left( \hbar \epsilon_{0} c\right)$. We also define a density distribution function as $\tilde{\rho}(z)$ = $\tilde{n}(z)/\mathcal{N}$, where the averaged atomic density is given by
\begin{equation}
\mathcal{N}=\frac{{\mathrm{OD}} \Gamma}{ \bar{\beta}L}.
\end{equation}
In the experiment, we observe the average atomic density $\mathcal{N}$ increases with the elongation of atomic ensemble \cite{Zhang2012}, as illustrated by the blue curve in Fig. S2.

Since the atomic cloud is located in the far-field region of the antenna horn, the incident microwave fields $\Omega_{\mathrm{M}}$ and $\Omega_{\mathrm{A}}$ are assumed to be the plane wave for the small medium. The equivalent absorption volume of atomic medium can be calculated by the integration of each absorption cross section in the $z$ distance. Note that the absorption cross section calculated here for microwave intensity $I_{\mathrm{M}}$ should be proportional to the square of atomic density, since the coupling constant for field amplitude is $\zeta_{\mathrm{M}}\propto \tilde{\rho}(z)$. Therefore, the averaged cross section of conversion medium is determined by
\begin{equation}
S_{\mathrm{M}}=\frac{1}{L} \int_{0}^{L}\pi r^{2}(z) \tilde{\rho}^2(z) \mathrm{~d} z,
\end{equation}
where
\begin{equation}
r(z)=\textit{w}_{\mathrm{C}} \sqrt{1+\left(\frac{z}{R}\right)^{2}},
\end{equation}
is the blue beam radius at the $z$ distance, and $R$ is the Rayleigh range of blue laser beam. In Eq. (S15), the averaged cross section $S_{\mathrm{M}}$ is a function of the ensemble length $L$ and the blue laser beam radius $\textit{w}_{\mathrm{C}}$. The errors in $L$, $\textit{w}_{\mathrm{C}}$ are independent and random, and thereby cause an uncertainty in $S_{\mathrm{M}}$ as follow:
\begin{equation}
\delta S_{M}=\sqrt{\left(\frac{\partial S_{M}}{\partial L} \delta L\right)^{2}+\left(\frac{\partial S_{M}}{\partial w_{C}} \delta w_{C}\right)^{2}}
\end{equation}
The radius $\textit{w}_{\mathrm{C}}$ is measured with a beam profile (Thorlabs BC106N-VIS) in a 5\% uncertainty. Together with the 0.1 mm ensemble uncertainty, the relative uncertainty of $S_{\mathrm{M}}$ in the optimized conversion is controlled within 4.7\%. Fig. S2 shows the numerical results of the mean radius $\overline{\textit{w}}=\sqrt{S_{\mathrm{M}}/\pi}$. In the optimized conversion, the optical depth is about 63 and the blue beam in the medium is expanded from a 54(3)~$\mu$m front end radius to a 79(1)~$\mu$m rear end radius, which corresponds to a mean radius of 66.0(1.5)~$\mu$m for the receiving cross section.

\section*{SIV. Microwave field strength calibration}

We calibrate the microwave field strength through the measurement of Rabi frequency $\Omega_\mathrm{M}$, as described previously in reference \cite{LiaoPRA2020}. As shown in Fig. S3(a), the frequency of field $\Omega_{\mathrm{S}}$ is red detuned by 100 MHz to drive the transition $|2\rangle\leftrightarrow|39D_{5 / 2}, m_{J}=1/2\rangle$ instead of $|2\rangle\leftrightarrow|39D_{3/2}, m_{J}=1/2\rangle$, while the frequency of probe laser is scanned around the two-photon resonance from -10 to 10 MHz within 100 $\mu$s using an acousto-optic modulator. In the four-level system represented in the subspace $\{|1\rangle,|2\rangle,|5\rangle,|4\rangle\}$, the atomic polarization corresponding to the $|1\rangle\leftrightarrow|2\rangle$ transition writes
\begin{eqnarray}
\begin{array}{ll}
&\varrho_{\mathrm{P}}(\Delta_{\mathrm{P}})=\frac{{\Omega_{\mathrm{P}}}(d_{4}d_{5}-\Omega_{\mathrm{M}}^{2})} {d_{2}d_{4}d_{5}-d_{2}\Omega_{\mathrm{M}}^{2}-d_{4}\Omega_{\mathrm{S}}^{2}},
\end{array}
\end{eqnarray}
where the complex detunings are $d_2=\Delta_{\mathrm{P}}-i \Gamma/2$, $d_4= \Delta_{\mathrm{4}}-i (\gamma_{4}+\gamma^{\prime}_{4}/2)$, and $d_5=\Delta_{\mathrm{5}}-i (\gamma_{5}+\gamma^{\prime}_{5}/2)$.
\begin{figure}[th]
\begin{center}
\includegraphics[width=8.3cm]{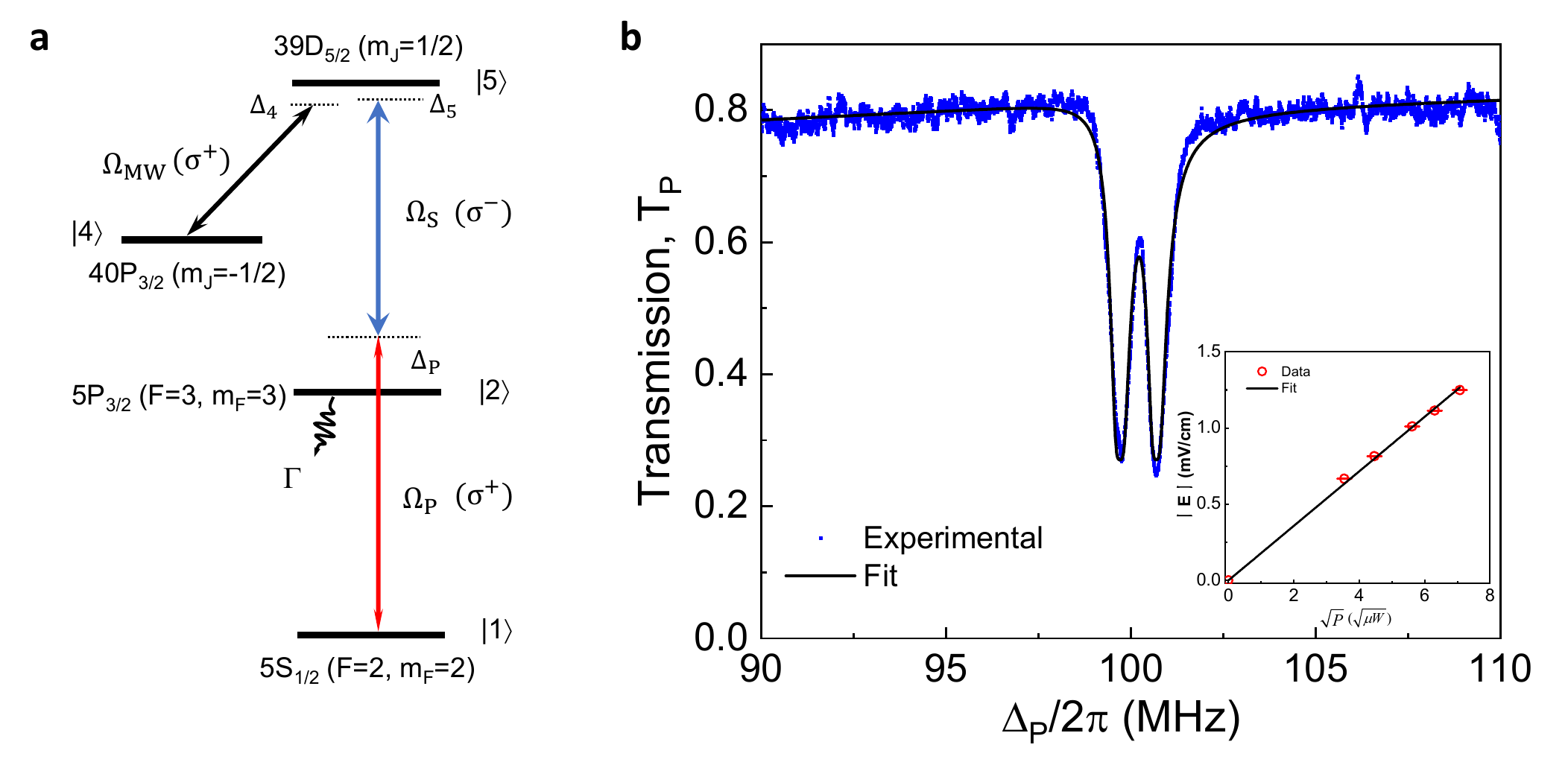}
\caption{\label{MW} \textbf{ Four-level EIA scheme for measuring the microwave electric-field amplitude. }
\textbf{a}, Relevant energy levels of $^{87}$Rb atoms. A weak probe laser, a strong coupling laser, and a microwave field are coupled to three electric dipole transitions $|1\rangle\leftrightarrow|2\rangle$, $|2\rangle\leftrightarrow|5\rangle$, and $|5\rangle\leftrightarrow|4\rangle$, respectively. \textbf{b}, A sample spectrum of probe transmission. The blue dots are experimental data averaged over 1000 cycles, and black line is the result of four-level susceptibility fitting. The inset displays the microwave field amplitude $|E|=\hbar \Omega_{\mathrm{M}} /|\boldsymbol{d}_{45}|$ against the microwave power input to the horn antenna, corresponding to a microwave intensity $I_{\mathrm{M}}=\frac{|E|^{2}}{2}\sqrt{\frac{\epsilon_{0}}{\mu_{0}}}$. The error bars are standard deviation of 60 measurements.
 }
\end{center}
\end{figure}

First, the strong microwave electric field is measured with an uncertainty less than 1\% in the EIA Autler-Townes splitting regime \cite{LiaoPRA2020}, corresponding to the microwave intensities $I_{M}>$ 4 pWmm$^{-2}$ in the experiment. A typical example of four-level EIA spectra is shown in Fig. S3(b). In the fitting procedure, we extract the free parameters including microwave Rabi frequency $\Omega_{\mathrm{M}}$, Rydberg dephasing rates $\gamma_{4}$ and $\gamma_{5}$. Note that this method also leads to a direct International System of Units traceable measurement of the microwave field amplitude\cite{GordonAPL2014}.
After that, we use these microwave intensities to calibrate the power meter reading from the microwave generator and then the calibrated power meter is used to extrapolate the microwave field at the low intensities $I_{M} <$ 4 pWmm$^{-2}$. Considering that the nonlinearity of the output power of microwave generator is less than 1\%, the total uncertainty of the quite low microwave intensities extrapolated from the Autler-Townes splitting regime would be less than 1.8\%. The inset of Fig. S3(b) shows the calibration data with regarding to the microwave field amplitude against microwave power input to the horn antenna. Furthermore, the calibration data shows this microwave electrometry operates with a repeatability better than 0.
\% in the Autler-Townes splitting regime. Similarly, the parameters $\Omega_{\mathrm{A}}$ and $\gamma_{3}$ are calibrated by using the four-level EIA scheme as described above.

Then we evaluate the overall uncertainty of conversion efficiency in regard to three quantities $P_{\mathrm{L}}$, $S_{\mathrm{M}}$ and $I_{\mathrm{M}}$ with uncertainties $\delta P_{L}$, $\delta S_{M}$ and $\delta I_{M}$. By using Eq.(2) in the main text, the efficiency uncertainty is determined by
\begin{equation}
\delta \eta=\sqrt{\left(\frac{\partial \eta}{\partial P_{L}} \delta P_{L}\right)^{2}+\left(\frac{\partial \eta}{\partial S_{M}} \delta S_{M}\right)^{2}+\left(\frac{\partial \eta}{\partial I_{M}} \delta I_{M}\right)^{2}}.
\end{equation}
The uncertainty calculation takes into account the optical power errors shown in Fig. 3a, the 1.8\% microwave intensity uncertainty, and the 4.7\% averaged cross section uncertainty.  The uncertainty $\delta \eta$ is smaller than 4.5\% for $I_{M}>$ 0.05 pWmm$^{-2}$, as shown in Fig. 3b. The large uncertainty for the data points in lower microwave intensity regime is primarily caused by the background noise fluctuation of the PMT detector.
We can derive with the data in Fig. 3b that the average conversion efficiency is $\eta = 82\%$ with a standard deviation $2\%$ and average uncertainty $7\%$.

\section*{SV. Propagation simulation for the converted field }

Finally, we provide a detail description of theoretical simulation presented in Fig. 2d of the main text, to illustrate the dependence of converted optical power on the optical depth. From the same waist along the propagation distance, the beam divergence of 780~nm laser is faster than that of 480~nm laser in the atomic ensemble. In the phase matching direction, the generated field $\Omega_\mathrm{L}$ that propagates near the edge of blue beam would spread out of the frequency-mixing volume, and thereby the atomic ensemble is acting as a two-level system for this fraction of generated field. We calculate the weighted mean of density matrix element for the converted field $\Omega_{\mathrm{L}}$, and its propagation equation writes
\begin{equation}
\left(\frac{1}{c} \partial_{t}+\partial_{z}\right) \Omega_{\mathrm{L}}=2 i \zeta_{\mathrm{L}} [\beta\varrho_{61}+(1-\beta)\frac{\Omega_{\mathrm{L}}}{-\Delta_{\mathrm{L}}-\frac{i\Gamma^{\prime}}{2}}].
\end{equation}
where $\beta$ denotes the ratio of the cross section of 480~nm beam over that of 780~nm beam at the propagation distance $z$, and $\zeta_{\mathrm{L}}$ is the optical coupling constant with respect to average density $\mathcal{N}$. For a given ensemble length $L$, average density $\mathcal{N}$ and input microwave Rabi frequency $\Omega_{\mathrm{M}}$, the conversion efficiency $\eta$ is simulated by using the Maxwell-Bloch equations, specifically, in which Eq. (S20) governs the propagation for the generated field $\Omega_{\mathrm{L}}$. In Fig. 2(d) of the main text, the fitted line of near-resonant scattering case is simulated with the parameters $\{\gamma_{3}, \gamma_{4}, \gamma_{5}, \Delta_\mathrm{P}, \Delta_{3}, \Delta_{4}, \Delta_{5}, \Delta_\mathrm{L} \}$ of $2\pi\times \{0.3, 0.2, 0.3, 0, 2.5, 0.9, 0, 0.6\}$ MHz, respectively. In the near-resonant conversion, the generated optical power in our setup is dissipated immensely by the resonant two-level system especially at the large optical depths. With the above average cross section $S_{\mathrm{M}}$ and microwave intensity $I_{\mathrm{M}}$, we therefore calculate the converted optical power $P_{\mathrm{L}}$, and both the off-resonant and all-resonant scattering simulation results agree well with the experimental data as shown in Fig 2(d) of the main text.

\section*{SVI. Thermal microwave background }
\begin{figure}[th]
\begin{center}
\includegraphics[width=7cm]{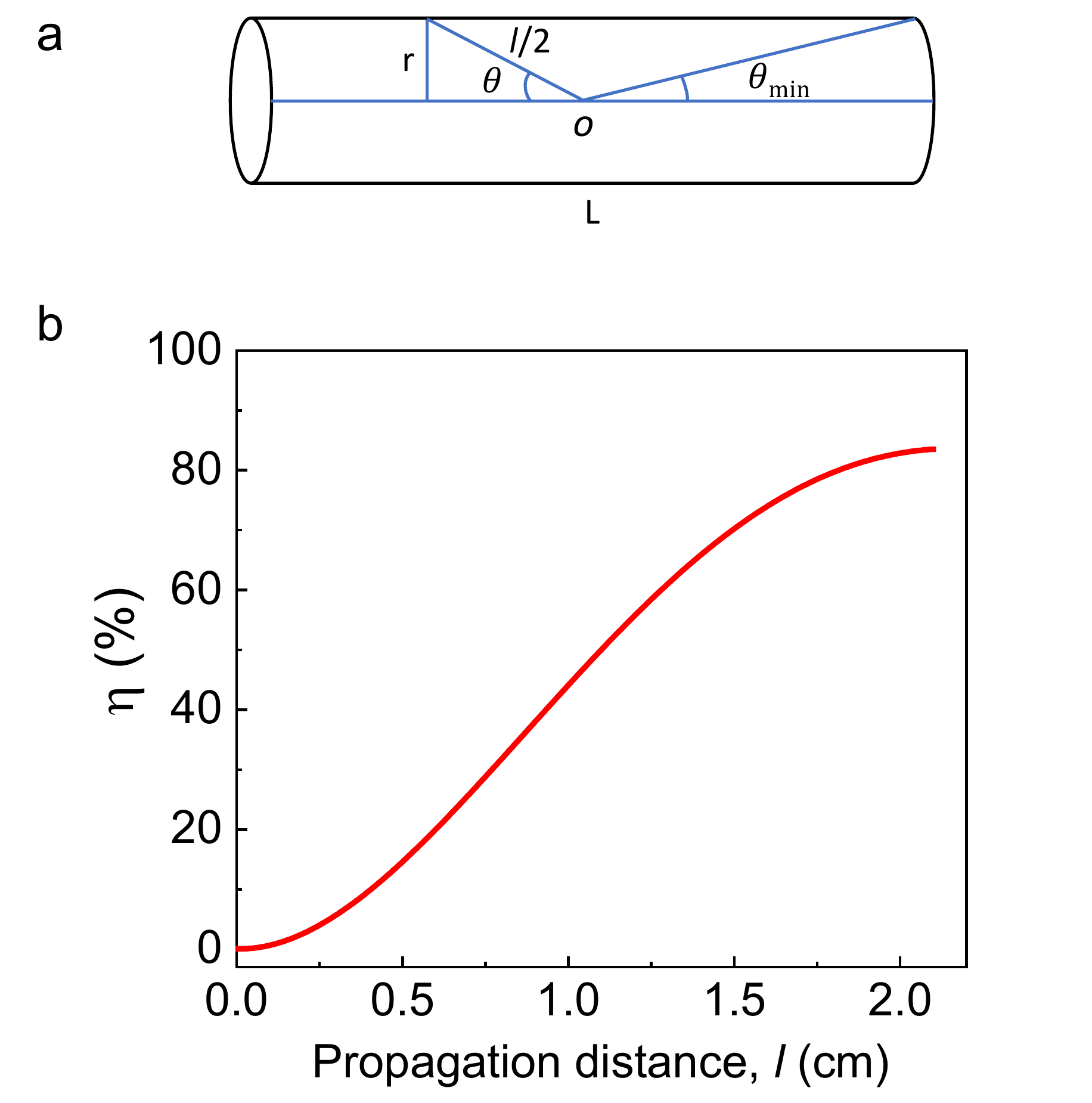}
\caption{\label{MW} \textbf{ Generation of optical photons due to thermal microwave background. }
\textbf{a}, The model of cylindric interaction medium. \textbf{b}, Conversion efficiency $\eta$ versus propagation distance $l$.
 }
\end{center}
\end{figure}

We here estimate the number of optical photons converted by the microwave background photons and find that the thermal photons contribute a fluctuation of about 2\% in the worst case. To calculate the thermal photon flux at 300 K microwave background, a basic assumption is that the background radiation is equivalent to a 300 K blackbody radiation. According to Stefan-Boltzmann law, energy flux density is given by
\begin{equation}
j=\int d \nu \int_{\Omega_{0}} d \Omega I(\nu, T) \cos \theta,
\end{equation}
where $d \Omega=\sin \theta d \theta d \varphi$ is the unit solid angle and $I(\nu, T)$ representing the microwave energy radiated by unit area of blackbody surface for unit time and unit solid angle at temperature $T$ and frequency of $\nu$ is given by
\begin{equation}
I(\nu, T)=\frac{2 h \nu^{3}}{c^{2}} \frac{1}{e^{h \nu / k T}-1}.
\end{equation}

In our system, the energy of microwave background photon within 1s is
\begin{equation}
Q=\int d \nu \int_{\Omega_{0}} d \Omega I(\nu, T) \cos \theta \cdot A.
\end{equation}
Here $A=2 \pi r^{2}+2 \pi r L$ is the surface area of cylindric interaction region where all the auxiliary lasers overlap in the atomic cloud, $L=2.1 $ cm is the length of the atomic cloud, and $r=66 \mu$m is the radius of beam waist. Then we obtain the radiation photon flux by dividing Eq. S(23) by $h \nu$ (energy of single photon)
\begin{equation}
\Phi(\nu, T)=\int d \nu \int_{0}^{2 \pi} d \varphi \int_{0}^{\pi / 2} d \theta \frac{2 \nu^{2}}{c^{2}} \frac{\cos \theta \sin \theta \cdot A}{e^{h \nu / k T}-1}
\end{equation}
At 300 K and 37 GHz, integrated over the 1 MHz conversion bandwidth, the radiation photon flux becomes $\Phi(\nu, T)\approx 141$ MHz. In our scheme, only the microwave $\Omega_{M}$ of $\sigma^{+}$ circular polarization can be up-converted. However, the blackbody radiation is unpolarized and isotropous \cite{LubinPRL1979}, so about half of the radiation photons satisfy $\sigma^{+}$ circular polarization. Within the interaction time T =10 $\mu$s, the number of radiation photons from all directions involving in the conversion process is given by
\begin{equation}
N=\frac{\Phi \times T}{2} \approx 705.
\end{equation}

For the needle-like interaction medium, the number of background photons from the radiation angle $\theta$ is expressed as
\begin{equation}
d S=N \frac{\int_{0}^{2 \pi}\frac {d \Omega}{d \varphi} d \varphi}{4 \pi}=N \frac{\sin \theta d \theta}{2}
\end{equation}
As shown in Fig. S4(a), we set the origin $O$ as the center of the interaction medium, and the effective propagation distance $l$ for the background photons at radiation angle $\theta$ is written  as
\begin{equation}
l=\frac{2 r}{\sin \theta}, \quad \theta \epsilon\left[\theta_{\min }, \pi-\theta_{\min }\right]
\end{equation}
where $\theta_{\min }=\arcsin \left(\frac{2 r}{L}\right)$; and, $l \cong L$ for the paraxial background photons radiated at the end surface of interaction volume.

The conversion efficiency decreases as the decrease of the effective propagation distance of radiation photons. We calculate the conversion efficiency $\eta$ against the propagation distance $l$ with the same parameters as that in experiment and the results are plotted in Fig. S4(b). By using Eq. S(27), only background microwave within a solid angle of 0.11 radians can be converted with an efficiency greater than 0.5 \%.

Considering the influence of conversion efficiency, the number of optical photons converted by the microwave background photons is given by
\begin{equation}
\begin{aligned}
S_{photon} &=\int_{\theta_{\min }}^{\pi-\theta_{\min }} \eta(l) d S+\frac{2 \eta_{\max}N \int_{0}^{\theta \min } \sin \theta d \theta}{4 \pi} \\
&=\int_{\theta_{\min }}^{\pi-\theta_{\min }} \frac{\eta(\theta) N \sin \theta d \theta}{2}+\frac{2 \eta_{\max } N \int_{0}^{\theta_{\min}} \sin \theta d \theta}{4 \pi} \\
&\approx 0.79,
\end{aligned}
\end{equation}
where $\eta_{\max}$ = 82\% is the highest conversion efficiency in the text. Therefore, only less than one optical photon is generated due to the background radiation. The main reason is that the geometry of the interaction medium is a needle-like (r = 66$\mu$m and L =2.1 cm), so the radiation solid angle that is applicable to efficient up-conversion is extremely small. Compared with the weakest field conversion which generates about 40 optical photons (as shown in Figure 3b), the thermal photons contribute a fluctuation of about 2\% in the worst case, but the total fluctuation is around 38\% primarily due to the technical noises. Hence, the effect of the thermal microwave background is non-resolvable in our experiment, and it does not affect the stated conversion efficiency.

\end{document}